\documentstyle[preprint,prc,eqsecnum,aps]{revtex}
\begin{document}
\tolerance=10000
\hbadness=10000
\vbadness=10000
\draft
\tighten
\preprint{\ }
\title{
     Impact Parameter Dependence of Multiple Lepton-Pair\\
     Production from Electromagnetic Fields }

\author{
     M. C. G\"u\c cl\"u$^{1,2}$ , J. C. Wells\thanks{\em
     New address:\ Harvard-Smithsonian Center for
     Astrophysics, Cambridge, MA 02138}$^{1,2}$,
     A. S. Umar$^{1,2}$, M. R. Strayer$^{1}$, and D. J. Ernst$^{1,2}$}
\address{\ \\ \ \\
 {$^1$Center for Computationally Intensive Physics, Physics Division,}\\
 {Oak Ridge National Laboratory, Oak Ridge, Tennessee 37831-6373, USA}\\
 {$^2$Vanderbilt University, Department of Physics and Astronomy,}\\
 {Nashville, Tennessee 37235, USA
\ \\
\ \\}
}
\date{\today}
\maketitle
\begin{abstract}
In relativistic heavy-ion collisions, the strong Lorentz-contracted
electromagnetic fields are capable of producing copious numbers of
lepton pairs through the two-photon mechanism. Monte Carlo techniques have
been developed that allow the exact calculation of production by this
mechanism when a semi-classical approximation is made to the motion of the
two ions. Here, we develop a hybrid Monte Carlo technique that enables us
to calculate the impact parameter dependence of the two-photon mechanism for
lepton-pair production, and by using this result, we obtain the probability
distribution for multiple-pair production as a function of impact parameter.
Computations are performed for S$+$Au and Pb$+$Pb systems at 200 A GeV and
160 A GeV, respectively. We also compare our results with the equivalent
photon approximation and elucidate the differences.
\end{abstract}
\ \\
\ \\
\pacs{25.75.+r, 34., 12.20.Ds, 02.70.Jn}
\narrowtext

\section{INTRODUCTION}
\label{sec:level1}

The proposed new colliding-beam accelerators -- designed to investigate
nuclear matter at high temperatures and densities by accelerating highly
charged heavy ions at fixed-target energies per nucleon up to $20$ TeV --
have motivated great interest during the last decade concerning
possible new electromagnetic phenomena.
When heavy ions collide at relativistic velocities, the Lorentz-contracted
electromagnetic fields in the space-time region near the collision are
sufficiently intense to produce large numbers of electron-positron pairs,
muon pairs, vector bosons, and possibly the yet-unconfirmed Higgs boson.
All these processes occur at nearly atomic distance scales
\cite{BS3,CBMS,BSPG}.
The phenomena involved are pervasive, impinging upon atomic, nuclear, and
particle physics.  The electromagnetic fields associated with these
collisions are intense as they are proportional to the Lorentz factor $\gamma$,
which is approximately the beam kinetic-energy in units of GeV per nucleon,
and the charge of the ion $Z$.
These parameters, together with the impact parameter, $b$, determine
the fields available in a collision.
The feature of electromagnetic particle production by heavy ions,
which piqued early studies, is the $Z^4$ enhancement through the coherent
action of all the constituent charges of the colliding partners.
The pulsed fields are strongly time dependent with a width of $b/\gamma$
and, thus, contain large Fourier components which give rise to relatively
large particle production probabilities.
Coherent particle production is most clearly
distinguished in peripheral collisions.

One of the most interesting of these processes,
from the perspective of both fundamental and practical importance,
is the sparking of the vacuum to produce electron-positron pairs.
Much study has been devoted to the problem of electron-positron pair
production during recent years in anticipation of
new experimental opportunities at the Relativistic Heavy-Ion
Collider (RHIC), currently under construction at
Brookhaven National Laboratory, as the immense predicted
fluxes of electrons cannot be ignored in detector or
accelerator design.
This facility will provide extremely relativistic
colliding beams of fully stripped ions as heavy as gold,
fully exposing the large charge of the atomic nucleus.

The primary physics goal of the RHIC project is the creation
and study of a so-called quark-gluon plasma.
This unique form of matter is expected to be formed in the central,
or near-central, collisions of heavy ions at extreme relativistic
energies \cite{Blaizot}.
The thermodynamic conditions attained in these central collisions
are expected to be such that the constituent quarks and gluons
of baryons and mesons become deconfined in a new and short-lived
plasma state.
Electron- and muon-pair production from hadronic interactions
have been widely discussed as a possible tool to
help probe the formation and the decay of the quark-gluon
plasma phase of matter \cite{KK86a,KK86b}.
In the conditions of such central collisions, lepton-hadron final-state
interactions are usually small, and, hence, the leptons carry direct
information on the space-time region of creation.
However, suggestions by several authors indicate that other sources
of lepton pairs might possibly mask the leptonic signals
originating from the plasma phase \cite{HG85,BS87b,Te86}.
Electromagnetic production from the vacuum of single- and multiple-lepton pairs
is a major contribution to this physical background
\cite{BS87c,BS3} and, therefore, must be understood in detail.
In addition, two very abundant electromagnetic processes constitute
the primary limitation to the lifetime of stored beams at RHIC.
One is a nuclear decay following the electromagnetic excitation of the
giant dipole resonance, and the second is the creation of an
electron-positron pair accompanied by the capture of the electron
in an atomic bound state of a participant heavy ion. Both processes
result in a change in the charge-to-mass ratio of the ion in the
storage ring causing it to be deflected out of the beam.

 From a fundamental perspective,
relativistic heavy-ion collisions provide an opportunity to study
nonperturbative quantum electrodynamics (QED) in an entirely
new and continuously varying energy regime, using an interaction
which is completely known due to the combination of very
high collision energies and electric charges.
The coherent, electromagnetic production of electron-positron pairs using
heavy ions is fundamentally different from other production mechanisms using
light particles at high energies, since in the former, the coupling
constant is strongly enhanced due to the large charge.

Nonetheless, perturbative methods are useful in studying
the problem of lepton-pair production, and reliable lowest-order
perturbative calculations have been used as input into design models for
RHIC \cite{CBMS,RBL90}.
Beginning with quantum-field theory, we derive a classical-field
method based on Feynman perturbation theory in the limit that
the momentum transfer of the photons is much smaller than the
momentum carried by the nuclei \cite{CBMS,Wu90}.
This is manifestly true for the coherent production of particles
through two-photon processes in relativistic heavy-ion collisions.
These lowest-order diagrams (see Fig.~\ref{figex}),
coupling lepton fields to classical
electromagnetic fields, have been evaluated exactly
using Monte Carlo \cite{CBMS,Ver83} and analytical techniques
\cite{Eby91,HTB94}, and predict cross sections consistent with experiments
for free electron-positron production performed by two independent
groups using collisions of S + Au \cite{Vane92} and S + Pt \cite{Ceres},
respectively, at fixed-target energies of 200 A GeV.
Furthermore, no indication of nonperturbative effects is observed in these
experiments using the relatively light sulfur projectile.
However, applying lowest-order perturbation theory to the production
of electron-positron pairs with heavy ions at high energies and small impact
parameters results in probabilities and cross sections which violate various
theoretical bounds such as unitarity \cite{ws90,CBMS}.
It is therefore clear that low-order perturbative calculations alone
are not adequate for smaller impact parameters at RHIC energies,
as higher-order damping effects must be included (e.g.\ the
creation of multiple pairs \cite{Baur90}).
Heavy ions have been used to produce electron-positron pairs
in collisions of U + Au at fixed-target energies of 0.96 A GeV,
and the observed cross sections are not in agreement with the low-order
calculations performed at these energies \cite{Bel93}.
New experiments using a variety of targets are scheduled for November,
1994 \cite{Vane94} using lead beams at fixed-target energies of 160 A GeV,
and for 1996 \cite{Gould96} using gold beams at fixed-target
energies of 12 A GeV during which electron-pair multiplicities and other
nonperturbative features of these collisions will be observed.

The Monte Carlo method described in Ref.\ \cite{CBMS} for evaluating
the two-photon Feynman diagrams employs an analytic integration over
the impact parameter  and, thus, does not provide information on
the impact-parameter dependence of the cross section.
In this present work, we generalize these techniques in order to calculate
the impact-parameter dependence of the two-photon diagrams
which has implications for the description of the electromagnetic
background for hadronic interactions, the study of strong-field
effects (e.g.\ pair multiplicities), and the study of the validity
of the equivalent-photon approximation, as is discussed below.
Various possibilities for experimental study of the impact-parameter
dependence for free electron- and/or muon-pair production are
discussed in Ref.\ \cite{FN90}.

Concerning certain classes of
detectors designed for nuclear physics experiments at RHIC,
the total integrated cross section may not be the most relevant
quantity for studies of the electromagnetic background.
If the detector triggers on a limited range of small impact parameters,
the pair production probability for these impact parameters would be
more relevant \cite{ws90}.
Due to the nature of the electromagnetic interaction, the particle production
process occurs over a large range of impact parameters.
The larger impact parameters produce weaker, more perturbative, fields,
but these impact parameters are favored by the geometrical weight
factor in the cross-section integration.
An integration over all impact parameters also includes
contributions from impact parameters where the two nuclei would collide.
The assumption that the nuclei continue on straight-line paths
for these nonperipheral impact parameters is not accurate.
Thus, one would like to isolate the relatively small
impact parameters which are still larger than the grazing impact parameter
for the study of strong-field effects.

Recent progress has been made in understanding the nonperturbative
nature of the production of multiple free electron-positron
pairs \cite{Ba90b,RBW91,Best92,HTB94p}.
Using the usual assumptions of the classical, strong nature
of the electromagnetic field generated by the heavy ions and
omitting the final state interactions among the produced leptons,
one can express the
probability for producing $N-$pairs as a Poisson distribution
whose mean value is the probability for producing a single pair
in lowest-order perturbation theory.
Therefore, the central ingredient needed for describing
the multiplicity distribution of electron-positron pairs within this
approximation is the impact-parameter dependence of the two-photon mechanism.

Historically, the two-photon process has been modeled
through the equivalent-photon approximation, which takes advantage of the
close relation between the interaction produced by a relativistic charged
particle and those due to incident electromagnetic waves
\cite{CFW,EJW,SJB,FEL,RHD,DLO,RAHG,AG,GBCA}.
In the equivalent-photon approximation, the equivalent-photon flux
associated with a relativistic charged particle is obtained via a Fourier
decomposition of the electromagnetic interaction \cite{Ja75,BLP}.
Cross sections are obtained by folding the elementary, real
two-photon cross section for the pair-production process
with the equivalent-photon flux produced by each ion.
The pair-production cross section is easier to compute
using the equivalent-photon approximation than by calculating the two-photon
diagram, and the results for the total cross section
are reasonably accurate, provided the incident-particle
Lorentz factor $\gamma$ is much greater than one, and that the energy
transferred via the photon is much less than $\gamma$.
However, details of the differential cross sections, spectra, and
impact-parameter dependence differ, especially when complicated
numerical cuts in the coordinates are applied in order to
compare with experiments.
Among the shortcomings of this approximation is
an undetermined parameter which corresponds to the minimum
impact parameter or the maximum momentum transfer, which makes it
difficult to get specific results.
As such, the method loses applicability at impact parameters less
than the Compton wavelength of the lepton \cite{HTB94}, which is the region
of greatest interest for the study of nonperturbative effects.
Reference \cite{BB88} reviews the application of the equivalent-photon
method in relativistic heavy-ion collisions through 1988.

Section \ref{sec:level2} introduces the basic two-photon approach
for computing the single-pair production and the generalization of
these results for calculating multi-pair production cross sections.
In this section we also discuss the impact-parameter dependence
of these cross-sections. In Section \ref{sec:num} we discuss the details
of the numerical techniques used for obtaining impact-parameter-dependent
cross sections. Section \ref{sec:results} outlines the results
for S$+$Au and Pb$+$Pb heavy-ion reactions. The paper concludes
with the discussion of the results in Section \ref{sec:con}.

\section{FORMALISM}
\label{sec:level2}
\subsection{Single Pair Production}
The cross section for lepton-pair production in relativistic heavy-ion
collisions via the two-photon process has been derived in \cite{CBMS}
using a semiclassical least-action principle. The result is
equivalent to utilizing the leading-order Feynman diagram and taking
the classical limit \cite{Wu90,BSAE,AMD} on the relative motion of
the two ions. In this formalism, the source currents appear as arising from
Lorentz-boosted charge distributions. We follow the formalism of \cite{CBMS},
only here we develop a technique for calculating cross sections as a
function of the impact parameter, rather than the integral over all impact
parameters, as was done previously.
The lowest-order two-photon process is pictured in Fig.~\ref{figex}.

Nucleus 1 is the nucleus which (in the center of velocity frame) moves with
a velocity --$\beta $, and nucleus 2 (in Fig.~\ref{fig1}) moves with
velocity +$\beta $, both parallel to the z-axis.  Their trajectories are
taken to be straight lines separated by
an impact parameter $b$. Throughout this paper, we use a system of units
with $\hbar =c=m=e=1$ where $m$ is the electron's rest mass.
The semiclassical coupling of electrons to the
electromagnetic field is given by the Lagrangian density
\begin{eqnarray}
{\cal L}_{int}(x)=-\overline{\Psi}(x)\gamma_{\mu}\Psi (x) A^{\mu} (x)\;,
\end{eqnarray}
which only depends on the field variables via the classical four-potential
$A^{\mu}$, where
\begin{eqnarray}
A^{\mu}=A^{\mu}(1) + A^{\mu}(2)\;,
\end{eqnarray}
and in the momentum space, the nonzero components of the potential from
nucleus 1 are
\begin{eqnarray}
A^{0}(1) &= & -8\pi^{2}Z\gamma^{2}\frac{\delta(q_{0}-\beta
q_{z})}{q_{z}^{2}+\gamma^{2}(q_{x}^{2}+q_{y}^{2})}\exp \left [i{\bf
q_{\perp}}\cdot \frac{\bf b}{2} \right ] \;,\nonumber \\
A^{z}(1) &= & \beta A^{0}(1) .
\end{eqnarray}
The potentials from nucleus 2 can be obtained from (3)
by the substitutions ${\bf b} \rightarrow -{\bf b}, \beta \rightarrow
-\beta$.
If we assume that the heavy-ion motion can be localized along definite
impact parameters, we can write the total inclusive singles $\sigma_{s}$ and
pair $\sigma_{p}$ cross sections:
\begin{eqnarray}
\sigma_{s} &= &\int d^{2}b {\cal N}_{s}({\bf b}) \;,\nonumber \\
\sigma_{p} &= &\int d^{2}b {\cal N}_{p}({\bf b}) \;,
\end{eqnarray}
where ${\cal N}_{s}({\bf b})$ is the singles multiplicity and
${\cal N}_{p}({\bf b})$ is the pair multiplicity of produced electrons.
In the strong field limit the multiplicity of electrons represents the mean
number of electrons produced out of the vacuum. It has been shown that, in
the perturbative limit, the pair multiplicity is the same as the singles
multiplicity, and the inclusive pair cross section is $\sigma_{p}$ $\simeq$
$\sigma$,
\begin{eqnarray}
\sigma = \int d^{2}b \sum_{k>0}
\sum_{q<0}|<\chi_{k}^{(+)}|S|\chi_{q}^{(-)}>|^{2} \;,
\label{eq:sig}
\end{eqnarray}
where the summation over the states $k$ is restricted to those above
the Dirac sea, and the summation over the states $q$ is restricted to
those occupied in the Dirac sea. The transition matrix element for direct
Feynman diagrams in Eq.\ (\ref{eq:sig}) has been derived in \cite{CBMS} as
\begin{eqnarray}
<\chi_{k}^{(+)}|S|\chi_{q}^{(-)}> = \frac {i}{2 \beta} \int
\frac{d^{2}p_{\perp}}{(2 \pi)^{2}}\exp \left \{ i \left [ {\bf p}_{\perp} -
\left [ \frac{{\bf k}_{\perp} + {\bf q}_{\perp}}{2} \right ] \right ] \cdot
{\bf b} \right \} {\cal A}^{(+)}(k,q:{\bf p}_{\perp}) \;.
\end{eqnarray}

Including both the direct and crossed Feynman diagrams, the results for the
cross section, as a function of impact parameter, can be obtained as
\begin{eqnarray}
\frac{d \sigma}{d b} & = & \frac{1}{4\beta^{2}}
\sum_{\sigma_{k} \sigma_{q}} \int \frac{d^{3}k d^{3}q d^{2}p_{\perp}
d^{2}p_{\perp}^{\prime}}{(2\pi)^{9}} b J_{0}(b\, |{\bf p}_{\perp}-{\bf
p}_{\perp}^{\prime}|) \nonumber \\ & & \times \left( {\cal
A}^{(+)}(k,q:{\bf p}_{\perp})+
{\cal A}^{(-)}(k,q:{\bf k}_{\perp}+{\bf q}_{\perp}-{\bf p}_{\perp})\right
)\nonumber \\ & & \times \left( {\cal A}^{(+)}(k,q:{\bf
p}_{\perp}^{\prime})+ {\cal A}^{(-)}(k,q:{\bf k}_{\perp}+{\bf q}_{\perp}-{\bf
p}_{\perp}^{\prime})\right )^{*} \;,
\label{eq:dsig}
\end{eqnarray}
where $\bf k$ ($\bf q$) is the momentum of the produced lepton
(anti-lepton), and ${\cal A}^{(\pm )}(k,q;{\bf p}_\perp )$ are given by
\begin{eqnarray}
{\cal A}^{(+)}(k,q:{\bf p}_{\perp}) & = & F({\bf k}_{\perp}-{\bf p}_{\perp}:
\omega_{1})F({\bf p}_{\perp}-{\bf q}_{\perp}:\omega_{2}){\cal T}_{kq}(
{\bf p}_{\perp}:+\beta) \;,\nonumber \\
{\cal A}^{(-)}(k,q:{\bf p}_{\perp}) & = & F({\bf k}_{\perp}-{\bf p}_{\perp}:
\omega_{2})F({\bf p}_{\perp}-{\bf q}_{\perp}:\omega_{1}){\cal T}_{kq}(
{\bf p}_{\perp}:-\beta) \;.
\end{eqnarray}
The quantity $F({\bf q},\omega)$ is the scalar part of the electromagnetic
field of the moving heavy ions in momentum space
\begin{equation}
F({\bf q}:\omega)=\frac{4\pi Z \gamma^{2}\beta^{2}}{\omega^{2}+\beta^{2}
\gamma^{2}|{\bf q}|^{2}}G_{E}(q^{2})\,f_Z(q^2) \;,
\end{equation}
where $G_E(q^2)$ is the form factor of the nucleon and $f_Z(q^2)$ is the
form factor of the nucleus.
The frequencies $\omega_1$ and $\omega_2$ of the virtual photons are
fixed  by energy conservation at the vertex where the photon is absorbed
\begin{eqnarray}
\omega_{1} & = & \frac{E_{q}^{(-)}-E_{k}^{(+)}+\beta(q_{z}-k_{z})}{2}\;,
\nonumber \\
\omega_{2} & = & \frac{E_{q}^{(-)}-E_{k}^{(+)}-\beta(q_{z}-k_{z})}{2}\;.
\end{eqnarray}
The quantity  ${\cal T}$ contains the propagator of the intermediate lepton
and the matrix elements for the coupling of the photon to the
leptons
\begin{eqnarray}
{\cal T}_{kq}({\bf p}_{\perp}:\beta) & = & \sum_{s}\sum_{\sigma_{p}}
\left [ E_{p}^{(s)}- \left [ \frac{E_{k}^{(+)}+E_{q}^{(-)}}{2}\right ]
+\beta\left [ \frac{k_{z}-q_{z}}{2} \right ] \right ]^{-1} \nonumber \\ & &
\times <u_{\sigma_{k}}^{(+)}|(1-\beta\alpha_{z})|u_{\sigma_{p}}^{(s)}>
<u_{\sigma_{p}}^{(s)}|(1+\beta\alpha_{z})|u_{\sigma_{q}}^{(-)}> \;.
\end{eqnarray}
Here,  $|u_{\sigma_{q}}^{(\pm )}>$ is the usual Dirac spinor and $\alpha_z$
is the z-component of the Dirac matrices.
\subsection{Multiple Pair Production}

In this section we will provide the basic formulas used in our
calculation of multi-pair cross sections. Recently, a number of
differing approaches have been developed to compute the
multi-pair cross sections \cite{Ba90b,RBW91,Best92,HTB94p}, all
resulting in a Poisson distribution for the multi-pair
impact-parameter-dependent probability function.
The prevailing assumption
is that the heavy ions do not suffer any recoil or energy loss while
electron-positron pairs are produced from the vacuum, which
seems to be a reasonable assumption when the pair energy is compared
with that of the heavy ions. One implication of the above
assumption is the approximation of straight-line, heavy-ion trajectories
for all impact parameters. Although this may not be
correct for central impact parameters, its reliability for
cross-section predictions depends on the physical region
for producing pairs. Provided that most of the pair production
cross section stems from peripheral heavy-ion trajectories, the
use of this assumption seems to be safe.

It has also been recognized that in most lowest-order perturbative
calculations, pair production probability violates unitarity for
small impact parameters and extreme energies.
It has been suggested that the summation of the classes of diagrams
resulting from the independent pair approximation can be used to
restore unitarity to the lowest-order perturbation theory.
These have been discussed in Ref.\ \cite{Ba90b} using the
sudden and quasiboson approximations, in Ref.\ \cite{RBW91} who
obtained a similar result by a straightforward summation of the
diagrammatically defined series in the perturbational expansion, and
in Ref.\ \cite{Best92} using a nonperturbative treatment and
neglecting the interference terms.
The resulting multi-pair production probability is described by a
Poisson form
\begin{eqnarray}
P_{N}(b)=\frac{{\cal P}(b)^{N} \exp[-{\cal P}(b)]}{N!}\;,
\label{multip}
\end{eqnarray}
where $N$ denotes $N$-pair production and
\begin{eqnarray}
{\cal P}(b)= \sum_{k>0}\sum_{q<0}|<\chi_{k}^{(+)}|S|\chi_{q}^{(-)}>|^{2}\;,
\end{eqnarray}
is the lowest-order perturbation result for the pair-production probability.
$\cal P$(b) can also be interpreted as the average number of electrons
produced out of the vacuum.
The $N$-pair cross section $\sigma_{N pair}$ is obtained by integrating
the $N$-pair probability over the impact parameter $b$.
\begin{eqnarray}
\sigma_{N pair}= \int d^{2}b P_{N}(b)\;\;\;\;,\;\;\;\; N=1,2,\ldots
\end{eqnarray}
In the two-photon formalism, as discussed above,
the total one-pair cross section is given by
\begin{eqnarray}
\sigma=\int d^{2}b {\cal P}(b)\;.
\end{eqnarray}
Therefore, from the above equation, we can write $\cal P$ (b)
\begin{eqnarray}
{\cal P}(b)=\frac{1}{2 \pi b} \frac{d \sigma}{d b},
\end{eqnarray}
which can be used in Eq.~(\ref{multip}) to calculate $N$-pair cross sections.
Consequently, our task in this paper is to determine a well-behaved
impact-parameter-dependent cross section.
In terms of the multi-pair cross section, the total cross section
for producing any number of pairs is given by:
\begin{equation}
\sigma_{pair}=\sum_{N=1}\,\sigma_{N\,pair}\;.
\label{eq:ptot}
\end{equation}

\section{NUMERICAL TECHNIQUES}
\label{sec:num}

For the calculation of the total cross section,
the differential cross section given by
Eq.\ (\ref{eq:dsig}) can be integrated over all impact parameters
analytically (it produces a simple delta function) and yields the result for
the total cross section as given in Ref.\ \cite{CBMS}.
However, the impact-parameter-dependent cross section is numerically
much more difficult and evaded any computations until the present.
This is due to the fact that
the function $J_0(b\,|{\bf p}_\perp - {\bf p}'_\perp |)$ is a rapidly
oscillating function, especially for large $b$.
One can add an additional $d^2p'_\perp$ integration to the Monte Carlo
technique given in Ref.\ \cite{CBMS} and attempt to do the integration
in Eq.\ (\ref{eq:dsig}) directly.
We tried up to 100 million Monte Carlo points, and, although
accuracy for small impact parameters was acceptable, we have not been
able to obtain convergence for large impact parameters.

We propose the following technique for circumventing this difficulty. We
divide the integration according to:
\begin{eqnarray}
\frac{d\sigma}{d b} = \int_{0}^{\infty}d q\, q\, b\, J_{0}(q b){\cal
F}(q)\;,
\label{eq:oned}
\end{eqnarray}
where ${\cal F}(q)$ is given by a nine-dimensional integral
\begin{eqnarray}
{\cal F}(q) & = & \frac{\pi}{8
\beta^{2}}\sum_{\sigma_{k}}\sum_{\sigma_{q}}\int_{0}^{2\pi} d \phi_{q}
\int \frac{d k_{z} d q_{z} d^{2} k_{\perp} d^{2} K d^{2} Q}{(2\pi)^{10}}
\nonumber \\
& & \times \left\{ F \left [ \frac{1}{2}({\bf Q-q});\omega_{1}
\right ]\right. F \left [{\bf -K}; \omega_{2} \right ] {\cal T}_{kq} \left
[{\bf
k}_{\perp}- \frac{1}{2}({\bf Q-q});+\beta \right ] \nonumber \\
& & + \left. F \left [ \frac{1}{2}({\bf Q-q});\omega_{1}
\right ] F \left [{\bf -K}; \omega_{2} \right ] {\cal T}_{kq} \left [{\bf
k_{\perp}-K} ;-\beta \right ] \right\}\nonumber \\
& & \times \left\{ F \left [ \frac{1}{2}({\bf Q+q});\omega_{1}
\right ] F \left [{\bf -q-K}; \omega_{2} \right ] {\cal T}_{kq} \left
[{\bf k}_{\perp}- \frac{1}{2}({\bf Q+q});+\beta \right ]\right. \nonumber \\
& & + \left. F \left [ \frac{1}{2}({\bf Q+q});\omega_{1}
\right ] F \left [{\bf -q-K}; \omega_{2} \right ] {\cal T}_{kq} \left
[{\bf k_{\perp}+q-K};-\beta \right ] \right\}\;,
\label{eq:foq}
\end{eqnarray}
where we have changed variables to
\begin{eqnarray}
{\bf p}_{\perp} & = & {\bf k}_{\perp}-\frac{1}{2}({\bf Q-q}) \nonumber \\
{\bf q}_{\perp} & = & {\bf k}_{\perp}-\frac{1}{2}{\bf (Q-q)+K} \nonumber \\
{\bf p}_{\perp}^{\prime} & = & {\bf k}_{\perp}-\frac{1}{2}{\bf (Q+q)} \;.
\end{eqnarray}
For a fixed value of $q$, the Monte Carlo technique of Ref.\ \cite{CBMS}
can be generalized to calculate the nine-dimensional integral of Eq.\
(\ref{eq:foq}). The details are provided in the Appendix.

This procedure results in a function, ${\cal F}(q)$, which is a
relatively smooth function
of $q$. The number of Monte Carlo points used is chosen such that
there is approximately a five percent error in each point we
generate for ${\cal F}(q)$. We fit a smooth function to the ${\cal F}(q)$
generated by the Monte Carlo integration and then perform the $q$ integral
analytically.
We check the procedure by integrating over the impact parameter and find
that we are always within the five percent error we set for ourselves in
comparison with the total cross section calculated using the methods of Ref.\
\cite{CBMS}.

\section{RESULTS}
\label{sec:results}

To demonstrate the ability of the technique presented here to produce
the impact parameter dependence of the two-photon electron-positron pair
production cross sections, we present results for two typical cases.
These are for electron-positron pairs produced in  S + Au  collision at
200 A GeV per nucleon, and in Pb + Pb collision at 160 A GeV per
nucleon in the fixed-target frame.
These two systems are studied because they correspond
to experiments either completed or planned for the immediate
future at CERN \cite{Vane92,Vane94}.
In all cases we set the form factor of the nucleon equal to 1 and use
a Woods-Saxon form factor for the nucleus. The use of regular
form factors for the nucleon, or uniform or Gaussian form factors for the
nucleus does not alter the results at these energies.

In Fig.~\ref{fig3}  we present the function
${\cal F}(q)$ from both the Monte Carlo calculation and from our smooth
fit. The smooth functions fit to ${\cal F}(q)$ are given by, for Pb + Pb
and S + Au  cases, respectively
\begin{eqnarray}
{\cal F}(q) & = & 3835.7 \;e^{-1.34029 q} , \nonumber \\
{\cal F}(q) & = & 163.5 \;e^{-1.3519 q}\;.
\label{eq:fit}
\end{eqnarray}
The differential pair production $d\sigma /db$ is then calculated by
integrating the smooth fit to ${\cal F}(q)$ according to Eq.\ (\ref{eq:oned}).
The results are presented in Fig.~\ref{fig4}
for S + Au  and Pb + Pb  cases, respectively. We see a smooth
exponential fall for large $b$ as expected. We may further integrate
these results over impact parameter. In Fig.~\ref{fig5}
we show the integrated cross section when integrated to a maximum
impact parameter, $b_{max}$, given by
\begin{eqnarray}
\sigma & = & \int_{0}^{\infty}d q\,q \int_{0}^{b_{max}}db\,b\,
J_{0}(q b){\cal F}(q) \nonumber \\
\,\,\,\,\, & = & \int_{0}^{\infty}d q\, b_{max}\, J_{1}(q b_{max})
{\cal F}(q)
\label{eq:bmax}
\end{eqnarray}
which as $b$ goes to infinity reproduces the total cross section
obtained in Ref.\ \cite{CBMS} using the exact integration over $b$.
Analytically one could also show that ${\cal F}(q=0)$
converges for large $b_{max}$ to the total cross section of
\cite{CBMS}.  This is done by taking the $b_{max}\longrightarrow\infty$
limit of Eq. (\ref{eq:bmax}) as follows
\begin{eqnarray}
\sigma & = & \int_{0}^{\infty}d x\, J_{1}(x)\,{\cal F}(\frac{x}{b_{max}})
\,\,\, , \,\,x=q\,b_{max} \nonumber \\
\,\,\,\,\, & \approx & \lim_{b_{max} \rightarrow \infty} {\cal F}(0)
\int_{0}^{\infty}d x\, J_{1}(x) \nonumber \\
\,\,\,\,\, & = & {\cal F}(0)\;\; ,
\end{eqnarray}
where we have taken ${\cal F}$ out of the integral, since, in the range
of $x$ for which the integral converges, ${\cal F}$ is essentially
a constant with argument very close to zero. The same result can also
be obtained by actually setting $q=0$ in the definition of ${\cal F}(q)$
given by Eq. (\ref{eq:foq}), and realizing that the resulting expression
is the same as the total cross section expression given in Ref. \cite{CBMS}.
Numerically, we have
obtained this result in Eq.\ (\ref{eq:fit}) with a less than two percent error.

We compare our two-photon results in Figs. \ref{fig4} and \ref{fig5}
with results obtained using the equivalent photon approximation as
expressed in \cite{BB88,RBW91}.
Using this equivalent photon method, a simple expression for
${\cal P}(b)$ is given by the lowest-order perturbative result as
\begin{eqnarray}
{\cal P}(b) \simeq \frac{14}{9\pi^{2}}(Z^{2}\alpha^{2})^{2} \left [
\frac{\lambdabar_{C}}{b} \right ]^{2}ln^{2} \left [\frac{\gamma \delta
\lambdabar_{C}}{2b} \right ] + \Delta(Z)\, ,
\label{eq:bsp}
\end{eqnarray}
where $\gamma$ is approximately the energy of one ion measured in the
frame of the other ion ($\gamma=2\gamma^{2}_{col}-1)$,
$\delta$ is a constant that has a value of 0.681.
$\Delta(Z)$ is a Coulomb modification which is ignored here.
In this equation, for impact parameters smaller than $\lambdabar_{C}$,
${\cal P}(b)$ goes to infinity very rapidly; i.e., the equivalent-photon
formula diverges for very small impact parameters \cite{HTB94}.
In addition, impact parameters larger than $\gamma \delta
\lambdabar_{C}/2$  exceed the region of validity for this formula.
However, for valid impact parameters, i.e.\ $\gamma \delta \lambdabar_{C}/2
\geq b \geq \lambdabar_{C} $, the simple equivalent-photon method gives
quite reasonable results although it overestimates the contribution to
pair production from the two-photon process by approximately five percent
in both cases considered, with the largest disagreement coming at small
impact parameters.
In Fig.\ \ref{fig5}, the equivalent-photon probabilities have been
integrated in their region of validity, whereas the integration limits for the
Monte Carlo evaluation of the two-photon diagrams are from zero to infinity.
We note that other formulations of the equivalent-photon approximation
may implement slightly different approximations from the ones used
to obtain Eq.\ (\ref{eq:bsp}) \cite{Vid93}, and, as a result, may produce
results somewhat different than the simple formula used here.

Taking advantage of our demonstrated ability to
compute the impact-parameter dependence of the two-photon
mechanism and the simple Poissonian form for the impact-parameter
dependence of the multiple-pair cross section (Fig.\ (\ref{multip})),
we compute the probability distributions for $N$-pair
production as a function of impact parameter $b$ for S + Au
(200 A GeV) and Pb + Pb (160 A GeV) collisions.
Our results show that the $N$-pair production probability has a
finite value at impact parameter $b=0$ and is continuous everywhere.
In Fig.~\ref{fig6} and Fig.~\ref{fig7} we see that one-pair production
completely dominates both the S + Au  and Pb + Pb collisions.
The two- and three-pair probability distributions are small and,
as expected, decrease rapidly for impact parameter values larger than
the Compton wavelength, although for the Pb + Pb case the two-pair
cross section has an appreciable 95b value.
The other values for two- and three-pair cross sections are shown
in Figs.~\ref{fig6} and \ref{fig7}.

For comparison, we also plot in Fig.~\ref{fig6} and
Fig.~\ref{fig7} the multiple-pair, impact-parameter-dependent
probability distributions calculated using the equivalent-photon
approximation (Eq.\ (\ref{eq:bsp})).
However, the inability of the equivalent-photon approximation
to accurately approximate the two-photon mechanism at impact
parameters below one Compton wavelength, along with the general
overestimation of the total pair-production cross section
by Eq.\ (\ref{eq:bsp}), results in the equivalent-photon approximation
overestimating the N-pair multiplicities.
For the lead collision considered,  the equivalent-photon approximation
overestimates the 2-pair cross section by $40\%$ and
the 3-pair cross section by a factor of $2$.
Once again, the equivalent-photon approximation has been
integrated here only over the range of valid impact parameters.

In summary, we have found the total one-pair cross section for
S + Au to be 160 barns and Pb + Pb to be 3569 barns by integrating
the two-photon diagram and using the Poissonian distribution
to extract the cross section for one-pair to be produced.
The cross section for any number of pairs can be found by summing the
contributions from all multi-pair cross sections as given by Eq.~\ref{eq:ptot}.
Another interesting quantity is the total cross section for multi-pairs, i.e.\
excluding the single-pair cross section.
For the lead collision considered, this cross section
has an appreciable value of approximately 100 barns which is about
$ 3\%$ of the total cross section.

\section{CONCLUSIONS}
\label{sec:con}
We have found a technique which is capable of calculating the semiclassical
two-photon mechanism for lepton-pair production as a function of the impact
parameter.
This required a generalization of the previously developed
Monte Carlo technique in order to be able to handle the oscillating
Bessel function which occurs in Eq.\ (\ref{eq:dsig}).
We find that utilizing a Monte Carlo technique to generate the
remainder of the integral, ${\cal F} (q)$, as a function of $q$,
fitting a smooth function to this result, and then performing the
final integral utilizing the smooth fit function, produces a workable
hybrid Monte Carlo technique.

Using the simple results obtained recently in understanding
the nonperturbative nature of the production of multiple, free
electron-positron pairs \cite{Ba90b,RBW91,Best92,HTB94p},
we use the calculated impact-parameter dependence of the
two-photon diagrams as the central ingredient needed to
calculate the multi-pair production cross sections.
Our calculations show that, for the systems and energies considered,
single-pair production is the dominant part of the total production
cross section as expected, since, at these energies, the two-photon
mechanism does not violate unitarity, indicating that nonperturbative
effects remain relatively small.
For the Pb + Pb case we find a substantial two-pair cross section
of 95 barns, which may yield itself to measurement.
However, the impact-parameter dependence of the two- and three-pair
cross sections are limited to a few Compton wavelengths ($b<10\lambdabar_C$).
For S + Au  and Pb + Pb collisions, we have found that a large portion of
the cross section of $e^{-}e^{+}$ pair production resides in the
region b $>$ 2R (R $\simeq$ 6.8 fm for Au, Compton wavelength for
electron $\lambdabar_{C}$=386 fm); thus, for low invariant masses
we can eliminate much of the concern over the hadronic
debris.
The simple equivalent-photon approximation does a fairly good
job of predicting the total pair-production cross section,
but its utility decreases when multiple-pair cross sections
are desired due to its inability to accurately describe
the small impact-parameter collisions.

It appears that for the lepton-pair production problem,
classical field treatment and quantum field theory
have complementary roles. In the classical field treatment, we assumed that
the momentum transfer of the photons is much smaller
than the momentum possesed by the nuclei; therefore, this is in
good agreement with the quantum field theory where the
energy and momentum conservation is exactly satisfied at all vertices of
Feynman diagrams. In relativistic heavy-ion collisions, coherent particle
production via two-photon processes does allow us to make the same
approximation because the nuclear form factor constrains the virtual photon
momenta where such an approximation is particularly good. Quantum field
theory provides guidelines for the classical field treatment, and tells us
what is the classical counterpart, especially for nuclear systems with spin,
and when such a treatment is valid. Classical field treatment also produces
a clear physical picture for the classical trajectories of heavy ions which
can be tagged experimentally, and enables us to properly simulate the
dependence of particle production in impact parameter space.

With the advent of higher-energy heavy-ion colliders, the study of the
physics of the two-photon process emerges as an exciting new field. The
process may be used as a means of production of exotic particles -- perhaps
the study of non-perturbative effects in QED.
In all cases,
an understanding of the impact-parameter dependence of the production
is necessary, and it is useful to be able to work in regions that are beyond
the validity of the equivalent-photon approximation.

\acknowledgements

This research was sponsored in part by the U.S. Department of Energy
under contract No. DE-FG05-87ER40376 with Vanderbilt University, and under
contract No. DE-AC05-84OR21400 with Oak Ridge National Laboratory managed by
Martin Marietta Energy Systems. In addition, this research was
partially supported by the U.S. Department of Energy High Performance
Computing and Communications Program (HPCC) as the Quantum Structure of
Matter Grand Challenge project.
One of us (M.C. Guclu) is also partially supported by the Turkish
Government.
The numerical calculations were carried out on CRAY-2 supercomputers
at the National Magnetic Fusion Energy Computer Center,
Laurence Livermore National Laboratory.

\clearpage

\appendix{}
\section{Monte Carlo Integration}

The ${\cal F}$(q) function in Eq.\ (\ref{eq:foq}) is of
the form
\begin{eqnarray}
{\cal F}(q) = F_{0}\int f(x_{1},x_{2},...,x_{8},q)d x_{1} d x_{2}\cdot \cdot
\cdot d x_{8}\;,
\end{eqnarray}
where we denote these coordinates by the eight-dimensional vector
\begin{eqnarray}
{\bf x} = \{\xi,\eta,k,\phi_{q},\theta_{Q},\phi_{Q},\theta_{K},\phi_{K}\}\;.
\end{eqnarray}
These variables are related to the variables defined in Eq.\ (\ref{eq:foq}) by
\begin{eqnarray}
\left ( \begin{array}{c}
k_{z} \\ q_{z} \end{array} \right ) & = & \gamma\, e^{\xi} \left (
\begin{array}{c}
\cos\eta \\ \sin\eta \end{array} \right ) , \\
\left ( \begin{array}{c}
Q_{x} \\ Q_{y} \end{array} \right ) & = & a_{Q}\,\tan\theta_{Q} \left (
\begin{array}{c}
\cos\phi_{Q} \\ \sin\phi_{Q} \end{array} \right ) , \\
\left ( \begin{array}{c}
K_{x} \\ K_{y} \end{array} \right ) & = & a_{K}\,\tan\theta_{K} \left (
\begin{array}{c}
\cos\phi_{K} \\ \sin\phi_{K} \end{array} \right ) , \\
\left ( \begin{array}{c}
q_{x} \\ q_{y} \end{array} \right ) & = & q \left (
\begin{array}{c}
\cos\phi_{q} \\ \sin\phi_{q} \end{array} \right ) , \\
\left ( \begin{array}{c}
k_{x} \\ k_{y} \end{array} \right ) & = & \left (
\begin{array}{c}
k \\ 0 \end{array} \right ) .
\end{eqnarray}
There are only eight integrals because we find that upon changing
to the above variables, the remaining integration can be done by symmetry.
Here, $\eta, \phi_{Q}, \phi_{K}$, and $\phi_{q}$ lie between 0 and
$\pi$, while $\theta_{Q}$ and $\theta_{K}$ lie between 0 and $\pi$/2.
The variable $\xi$ is used to set the upper limit for $k_{z}$
and $q_{z}$. The scale factors $a_Q$ and $a_K$ are taken to be

\begin{eqnarray}
a_Q=\omega/\gamma \;,
a_K=\omega/\gamma \;.
\end{eqnarray}
Monte Carlo methods reduce the Eq.~(\ref{eq:foq}) to a summation
\begin{eqnarray}
{\cal F} (q) = \sum f(x_{i1},x_{i2},...,x_{i8},q) \Delta^{8}x \;,
\end{eqnarray}
where $\Delta^{n}x$ is the volume element of the subregion.
For a finite volume V, we divide the integral region to equal volume
elements and obtain
\begin{eqnarray}
{\cal F} (q) = \frac{V}{N} \sum f(x_{i1},x_{i2},...,x_{i8},q) \;
\end{eqnarray}
where N is the number of subvolume elements. We can also monitor the
fluctuation of the results, and according to the central limit theorem for
large values of N, we can write the associated error
\begin{eqnarray}
\delta^{2} = (<f^{2}> - <f>^{2})/N\;,
\end{eqnarray}
where
\begin{eqnarray}
<f^{2}> = \frac{1}{N}\sum f^{2}(x_{i1},x_{i2},...,x_{i8},q) \nonumber \\
<f> = \frac{1}{N}\sum f(x_{i1},x_{i2},...,x_{i8},q) \;.
\end{eqnarray}

\begin{figure}[htb]
\caption{Direct (a) and crossed
(b) Feynman diagrams for pair production in a heavy-ion collision.}
\label{figex}
\end{figure}
\begin{figure}[htb]
\caption{Schematic diagram depicting a relativistic heavy-ion collision.}
\label{fig1}
\end{figure}
\begin{figure}[htb]
\caption{The function $\cal F$(q)
versus q for the production of electron-positron pairs. The points are the
results of the Monte Carlo calculation and the smooth curve is our fit to
these points.}
\label{fig3}
\end{figure}
\begin{figure}[htb]
\hspace{0.75in}
\caption{ The differential cross section $d \sigma /db$ versus impact
parameter $b$ for electron-positron pair production.
Solid lines: exact
numerical result; dot-dashed lines: equivalent-photon approximation
(Ref.\ \protect\cite{RBW91}).}
\label{fig4}
\end{figure}
\begin{figure}[htb]
\hspace{0.75in}
\caption{
The integrated cross section as a
function of $b_{max}$ for electron-positron pair production.
Solid lines:exact
numerical result;dot-dashed lines:equivalent-photon approximation
(Ref.\ \protect\cite{RBW91}).}
\label{fig5}
\end{figure}
\begin{figure}[htb]
\hspace{0.75in}
\caption{
Probability distribution for $N$-pair production as a function
of impact parameter $b$ for Pb + Pb collision. Solid lines: exact
numerical result; dot-dashed lines: equivalent-photon approximation
(Ref.\ \protect \cite{RBW91}).
Total $N$-pair cross sections are given in units of barns.}
\label{fig6}
\end{figure}
\begin{figure}[htb]
\hspace{0.75in}
\caption{
Probability distribution for $N$-pair production as a function
of impact parameter $b$ for S + Au collision. Solid lines: exact
numerical result; dot-dashed lines: equivalent-photon approximation
(Ref.\ \protect\cite{RBW91}).
Total $N$-pair cross sections are given in units of barns.}
\label{fig7}
\end{figure}
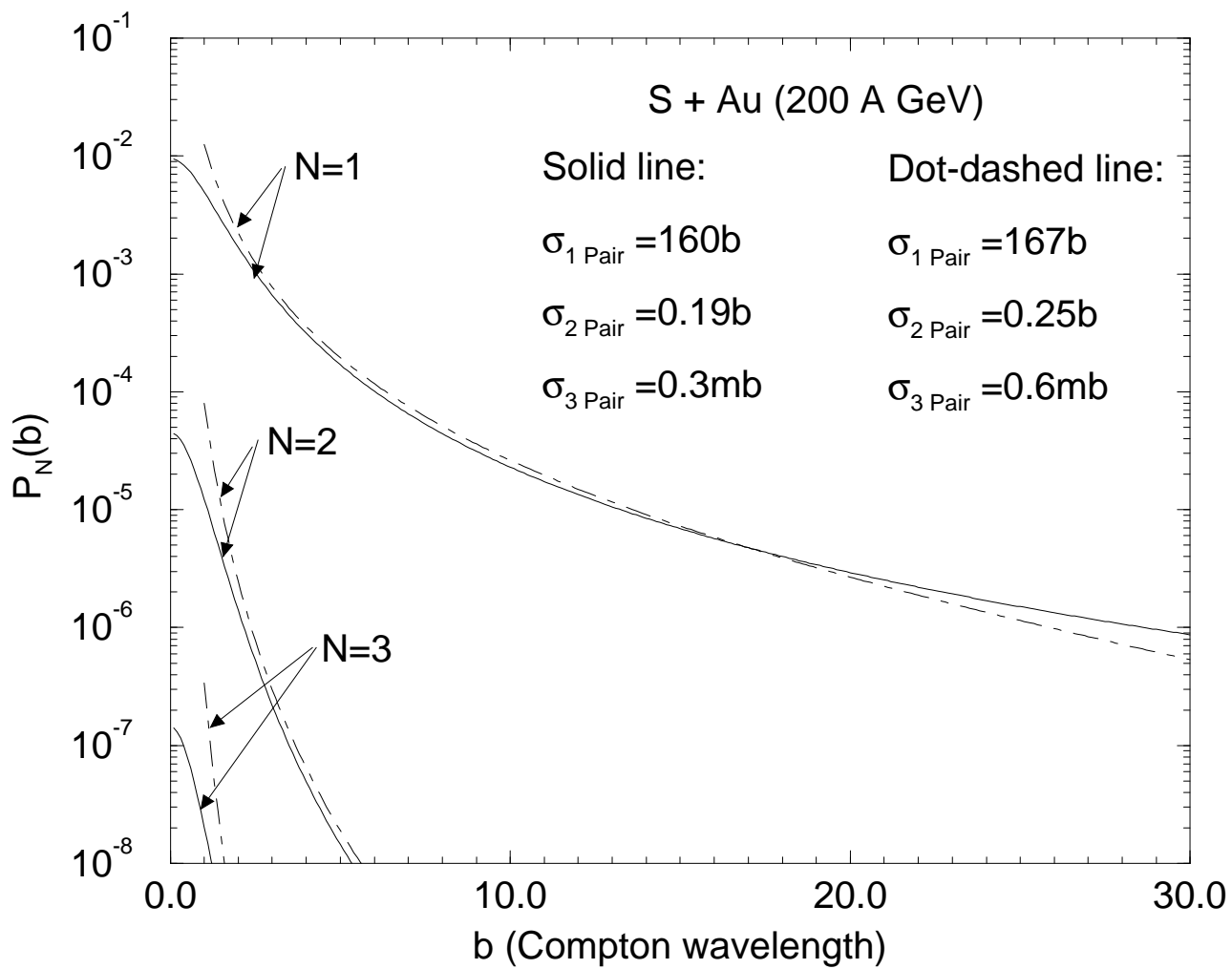
\end{document}